\documentclass[letterpaper, 10 pt, conference]{ieeeconf}
\IEEEoverridecommandlockouts                       
\let\labelindent\relax
\usepackage{enumitem}

\usepackage[utf8]{inputenc}
\usepackage[english]{babel}
\usepackage{multirow}
\usepackage{graphicx}
\graphicspath{ {./images/} }

\usepackage{cmap}

\usepackage[flushleft]{threeparttable}
\usepackage{svg}
\usepackage{booktabs}
\usepackage{graphicx}
\usepackage{tabularx}
\usepackage{etoolbox}
\usepackage{nomencl}
\usepackage[linesnumbered,ruled,vlined]{algorithm2e}
\usepackage{array}
\makenomenclature
\usepackage{amsfonts}
\usepackage{breqn}
\usepackage{color}
\usepackage{makecell}
\usepackage{amsmath}
\usepackage{cite}

 \usepackage{epstopdf} 

\usepackage{textcomp}
\usepackage{stfloats}
\usepackage{url}
\usepackage{verbatim}
\SetKwInput{KwInput}{Input}                
\SetKwInput{KwOutput}{Output}              
\usepackage{hyphenat}
\usepackage{hyperref}
\hypersetup{
    colorlinks=false,
    linkcolor=blue,
    filecolor=magenta,      
    urlcolor=cyan,
    pdftitle={Overleaf Example},
    pdfpagemode=FullScreen,
    }

\usepackage{comment}
\usepackage{subcaption}
\newif\ifarxiv
\arxivtrue

\renewcommand{\baselinestretch}{0.999}

\title{\LARGE \bf Restoring AC Power Flow Feasibility from\\ Relaxed and Approximated Optimal Power Flow Models}

\author{Babak Taheri and Daniel K. Molzahn
\thanks{B. Taheri and D.K. Molzahn are with the School of Electrical and Computer Engineering,
        Georgia Institute of Technology. \{taheri, molzahn\}@gatech.edu.
        Support from NSF EPCN award \#2145564.%
        }
}

\begin{document}
\maketitle

\begin{abstract}

\ifarxiv
To address computational challenges associated with power flow nonconvexities, significant research efforts over the last decade have developed convex relaxations and approximations of optimal power flow (OPF) problems. However, benefits associated with the convexity of these relaxations and approximations can have tradeoffs in terms of solution accuracy since they may yield voltage phasors that are inconsistent with the power injections and line flows, limiting their usefulness for some applications.
\else
\fi
Inspired by state estimation (SE) techniques, this paper proposes a new method for obtaining an AC power flow feasible point from the solution to a relaxed or approximated optimal power flow (OPF) problem. By treating the inconsistent voltage phasors, power injections, and line flows analogously to noisy measurements in a state estimation algorithm, the proposed method yields power injections and voltage phasors that are feasible with respect to the AC power flow equations while incorporating information from many quantities in the solution to a relaxed or approximated OPF problem. We improve this method by adjusting weighting terms with \textcolor{black}{an approach inspired by algorithms used to train machine learning models}. We demonstrate the proposed method using several relaxations and approximations. The results show up to several orders of magnitude improvement in accuracy over traditional methods.
\end{abstract}




\section{Introduction}
\label{sec:Introduction}
Optimal power flow (OPF) problems are central to many tasks in power systems. OPF problems optimize an objective, e.g., generation cost, while satisfying inequalities on the generators' power outputs, voltage magnitudes, line flows, etc. as well as equality constraints from a power flow model. 

The AC power flow equations accurately model the steady-state behavior of a power system by relating the complex power injections and line flows to the voltage phasors. OPF problems that use an AC power flow model are nonconvex and NP-Hard%
\ifarxiv
~\cite{bukhsh2013,molzahn2017,bienstock2019strong}.
\fi
\ifarxiv
The computational challenges from power flow nonconvexities are further compounded when solving OPF problems that consider, e.g., discreteness and uncertainty~\cite{barrows2014,roald2022review}.
\else
\fi
To address these challenges, OPF problems are often simplified using convex relaxations and approximations of the AC power flow equations. Common relaxations and approximations yield semidefinite programming (SDP)~\cite{lavaei2011zero}, second-order cone programming (SOCP)~\cite{jabr2006radial,coffrin2015qc}, and linear programming~\cite{coffrin2014linear} formulations; see~\cite{molzahn2019} for a survey.

Convex relaxations can certify problem infeasibility, obtain bounds on the optimal objective values, and yield globally optimal decision variables when the relaxation is tight \cite{molzahn2019}. When applied appropriately, power flow approximations can also provide useful results.
\ifarxiv
Accordingly, power flow relaxations and approximations are frequently used to convexify power systems optimization problems; see, e.g.,~\cite{roald2022review,molzahn2019,venzke2017,bestuzheva2020}. 
\else
\fi

The benefits of power flow relaxations and approximations can come at the expense of accuracy since the results may be infeasible or suboptimal with respect to the nonconvex OPF problem. This occurs due to inaccuracies in the power flow model, i.e., the voltage phasors from the solution to the relaxed or approximated OPF problem may be inconsistent with the power injections and line flows. This is problematic for many practical applications that require a solution which satisfies the power flow equations. Prior work includes conditions which guarantee that some relaxations will always provide globally optimal decision variables~\cite{molzahn2019,Low2014_exactness}, but these conditions are limited to special classes of problems that are not relevant to many practical settings.


Solution inaccuracies motivate the development of methods to restore voltage phasors and power injections that satisfy the AC power flow equations from relaxed or approximated power flow models. There are three types of methods in the literature for this purpose. The first type adds penalty terms to the objective function of a relaxed OPF problem; see, e.g.,~\cite{madani2014convex}. Appropriate choice of penalty terms can result in the relaxation being tight for the penalized problem, yielding feasible and nearly optimal solutions to some OPF problems. However, determining appropriate penalty parameters can be difficult and is often done in an ad hoc fashion that may require multiple time-consuming evaluations of the relaxed or approximated problem~\cite{venzke2020inexact}. 
The second type iteratively updates power flow relaxations and approximations within an algorithm that seeks a local optimum; see, e.g.,~\cite{tian2019recover} \textcolor{black}{which uses a difference of convex programming approach}. 
\textcolor{black}{Reference~\cite{fang2021ac} proposes another method of this second type. Starting with an infeasible point obtained using a power flow approximation, the goal of~\cite{fang2021ac} is to make small modifications to the outputs of a few generators to obtain a operating point that satisfies both the equality and inequality constraints of an OPF problem.} 
Methods of this second type may find high-quality operating points, but require good initializations and possibly involve repeated evaluations of the relaxed or approximated problem, which can be computationally expensive. See~\cite[Ch.~6]{molzahn2019} for a survey of these first two types of methods.

To avoid the challenges inherent to these first two types of methods, the third type uses a more straightforward and faster approach. Specifically, these methods simply fix certain values from the solution to a relaxed or approximated problem and then solve a power flow problem to obtain values for the remaining variables that are consistent with the AC power flow equations. There are multiple approaches for formulating these power flow problems. For instance, one method simply fixes the active power injections and voltage magnitudes at (non-slack) generator buses to formulate a power flow problem that is solved with traditional Newton-based methods~\cite{venzke2020inexact}. Alternatively, one could solve the power flow problem resulting from fixing the active and reactive power injections at (non-slack) generator buses, or one could substitute the voltage magnitudes and angles from the relaxed or approximated solutions into the power flow equations to get consistent values for active and reactive power generation.
Crucially, this third type of method may lead to values for some variables that do not match the solution to the relaxed or approximated problem. Moreover, the resulting point may violate the OPF problem's inequality constraints. Nevertheless, it is often important to get an AC power flow feasible point, ideally one that is as close as possible to the true OPF solution. The problem formulation, methodology, and numerical comparisons in this paper consider this third type of method.

Inspired by state estimation algorithms, we propose a new restoration method with significantly improved accuracy. Rather than fixing a subset of variables, we formulate an unconstrained optimization problem that considers information regarding voltage phasors, power injections, and line flows from the relaxed or approximated solution. Our method treats inconsistencies in these quantities due to power flow model inaccuracies analogously to measurement errors in state estimation. The use of more information from the relaxed or approximated solution enables restoration of a point that is closer to the true OPF solution while enabling us to borrow mathematical machinery from state estimation. 

The proposed method provides the freedom to choose weights that are analogous to the variances of the sensors' noise levels in state estimation algorithms. Rather than physically informed quantities obtained from sensor accuracies, the weights in our method are parameters that we can select based on the inconsistencies (with respect to the AC power flow equations) among various quantities in the solution to the relaxed or approximated OPF problem. This choice has a significant impact on the results. Since the inconsistencies in the solutions to the relaxation or approximation are not known a priori, choosing good values for these weights is challenging. We therefore determine these weights using \textcolor{black}{an approach inspired by algorithms used to train machine learning models}. During off-line computations, we solve a large number of OPF problems and their relaxations or approximations to get a training dataset. We then use a gradient descent algorithm to iteratively update the weights based on sensitivity information from the proposed restoration method. These updates drive the weights to values that minimize a loss function formulated as the squared difference between the true OPF solutions and restored points across the full set of training data. We then use the weights to recover the solutions during on-line calculations.
We demonstrate the proposed restoration method using various convex relaxations and approximations of OPF problems, with the results showing several orders of magnitude improvements in accuracy for some instances. 

To summarize, this paper's main contributions are:
\begin{itemize}[leftmargin=17pt]
    \item Developing a method inspired by state estimation techniques in order to restore AC power flow feasibility from the solutions to relaxed and approximated OPF problems.
    
    \item Improving this method by tuning weighting parameters via \textcolor{black}{an approach inspired by algorithms used to train machine learning models.}
    
    \item Demonstrating the performance of the proposed method using several power flow relaxations and approximations.
    
    
    
\end{itemize}

\textcolor{black}{In addition to the relaxations and approximations that are the subject of this paper, we note that there are also emerging machine learning approaches (e.g.,~\cite{pan2022deepopf, chatzos2022, Zamzam2020, kody2022}) that build more complicated models based on the results of many simulations. Similar to relaxations and approximations, the outputs of typical machine learning approaches require post-processing to obtain a point that satisfies the AC power flow equations. Thus, our approach is complementary to these emerging machine learning based approaches for solving OPF problems. Our ongoing work is assessing the performance of solution restoration approaches when applied to the outputs of various machine learning models.}

The remainder of this paper is organized as follows. Section~\ref{sec:Preliminaries}
describes the OPF problem and several convex relaxations and approximations. Section~\ref{sec:Operating Point Recovery Algorithm} presents the proposed operating point restoration method. Section~\ref{sec:Results and Discussion} demonstrates the proposed method with numerical experiments. Finally, Section~\ref{sec:Conclusion} concludes the paper and discusses future work.

\section{Preliminaries}
\label{sec:Preliminaries}
This section reviews the OPF problem along with the SDP, SOCP, and quadratic convex (QC) relaxations and the linear programming AC (LPAC) approximation. A survey of these and other relaxations and approximations is provided in~\cite{molzahn2019}.

\subsection{AC Optimal Power Flow}
We first introduce notation. Let $\mathcal{N}$ and $\mathcal{E}$ denote the sets of buses and lines, respectively. Each bus $i\in\mathcal{N}$ has a voltage phasor $V_i$, complex power demand $S^d_i$, shunt admittance $Y_i^S$, and complex power generation $S^g_i$. (Buses without generators are modeled with zero generation limits.) Complex power flows into each terminal for each line $(j,k)\in\mathcal{E}$ are denoted as $S_{jk}$ and $S_{kj}$. Each line $(j,k)\in\mathcal{E}$ has admittance parameters $Y_{jk}$ and $Y_{kj}$. Real and imaginary parts of a complex number are denoted as $\Re(\,\cdot\,)$ and $\Im(\,\cdot\,)$, $(\,\cdot\,)^*$ indicates the complex conjugate, and $(\,\cdot\,)^T$ indicates the transpose. Upper and lower bounds are denoted as $(\overline{\,\cdot\,})$ and $(\underline{\,\cdot\,})$. The OPF problem is:

\begin{subequations}
\label{eq:opf}
\begin{align}
& \label{eq:ac_pf1}\min \sum_{i \in N} c_{2 i}\left(\Re\left(S_{i}^{g}\right)\right)^{2}+c_{1 i} \Re\left(S_{i}^{g}\right)+c_{0 i} \hspace*{-10em} \\
\nonumber & \text{s.t.} ~~~ (\forall i\in\mathcal{N}, ~\forall (j,k)\in\mathcal{E})\\
& \label{eq:ac_pf8} \mathbf{W}_{jk}=V_j^{\vphantom{*}} V^{*}_k, ~ \mathbf{W}_{kj}=V_k^{\vphantom{*}} V^{*}_j, ~ \mathbf{W}_{ii}=V_i^{\vphantom{*}} V^{*}_i \\
& \label{eq:ac_pf9} \left(\underline{V}_{i}\right)^{2} \leq \mathbf{W}_{i i} \leq\left(\overline{V}_{i}\right)^{2} \\
& \label{eq:ac_pf10} \underline{S}^{g}_i \leq S^{g}_i \leq \overline{S}^{g}_i \\
& \label{eq:ac_pf11} \left|S_{jk}\right| \leq \overline{S}_{jk}, ~\left|S_{kj}\right| \leq \overline{S}_{jk}\\
& \label{eq:ac_pf12} S^{g}_i-S^{d}_i - \left(Y_i^S\right)^* \mathbf{W}_{ii}=\!\!\sum_{(i, j) \in \mathcal{E}}\!\! S_{i j} + \!\!\sum_{(k, i) \in \mathcal{E}}\!\! S_{ki}\\
& \label{eq:ac_pf13}   S_{jk}=Y_{jk}^{*} \mathbf{W}_{jj}-Y_{jk}^{*} \mathbf{W}_{jk} \\
& \label{eq:ac_pf14} S_{kj}=Y_{kj}^{*} \mathbf{W}_{kk}-Y_{kj}^{*} \mathbf{W}_{kj}^{*} \\
& \label{eq:ac_pf15} \tan \left(-\overline{\theta}_{jk}\right) \Im\left(\mathbf{W}_{jk}\right) \leq \Re\left(\mathbf{W}_{jk}\right) \leq \tan \left(\overline{\theta}_{jk}\right) \Im\left(\mathbf{W}_{jk}\right).
\end{align}
\end{subequations}
The OPF problem minimizes an objective function, here chosen to be generation cost in~\eqref{eq:ac_pf1} with quadratic coefficients $c_{2i}$, $c_{1i}$, and $c_{0i}$. The products of voltage phasors are collected in a Hermitian matrix $\mathbf{W}$ as described in~\eqref{eq:ac_pf8}. Voltage magnitude limits are imposed in \eqref{eq:ac_pf9}. The generators' output limits are enforced in \eqref{eq:ac_pf10} (interpreted as bounds on the active and reactive power outputs). Apparent power flow limits are specified in \eqref{eq:ac_pf11}. Constraint \eqref{eq:ac_pf12} enforces complex power balance at each bus and complex power flows are defined in \eqref{eq:ac_pf13} and \eqref{eq:ac_pf14} for each line. Limits on phase angle differences across lines are imposed in \eqref{eq:ac_pf15}. All nonconvexities in~\eqref{eq:opf} are associated with the products in~\eqref{eq:ac_pf8}. 


\subsection{Semidefinite Programming (SDP) Relaxation}
The nonconvex constraint \eqref{eq:ac_pf8} can be equivalently represented by the pair of constraints $\mathbf{W} \succeq 0$ (i.e., $\mathbf{W}$ is positive semidefinite) and $\textrm{rank}(\mathbf{W}) = 1$. Neglecting the rank constraint yields a semidefinite programming relaxation of the OPF problem~\cite{lavaei2011zero}:
\begin{equation} \label{eq:socp1}
\min \eqref{eq:ac_pf1} ~~ \textrm{s.t.} ~~ \eqref{eq:ac_pf9}\text{--}\eqref{eq:ac_pf15}, \mathbf{W} \succeq 0.
\end{equation}



\subsection{Second-Order Cone Programming (SOCP) Relaxation}
Rather than jointly considering all elements in the matrix $\mathbf{W}$, the SOCP relaxation separately convexifies individual elements from \eqref{eq:ac_pf8} using the fact that all positive semidefinite matrices have nonnegative principal minors~\cite{jabr2006radial}. Considering the principal minors associated with both the diagonal entries and the $2 \times 2$ submatrices yields the SOCP relaxation:
\begin{equation} \label{eq:socp2}
\min \eqref{eq:ac_pf1} ~~ \textrm{s.t.} ~~ \eqref{eq:ac_pf9}\text{--}\eqref{eq:ac_pf15}, |\mathbf{W}_{jk}|^{2} \leq W_{jj} W_{kk}, ~ \mathbf{W}_{ii} \geq 0.
\end{equation}



\subsection{Quadratic Convex (QC) Relaxation}
The QC relaxation described in~\cite{coffrin2015qc} introduces additional variables and constraints to better model phase angle relationships and current flows. Define new variables $v_i$, $\theta_i$, and $l_{jk}$ to represent the voltage magnitude and phase angle at each bus $i\in\mathcal{N}$ and the squared magnitude of the current flow on line $(j,k)\in\mathcal{E}$. Let $\left\langle\,\cdot\right\rangle^T$, $\left\langle\,\cdot\right\rangle^M$, $\left\langle\,\cdot\right\rangle^C$, and $\left\langle\,\cdot\right\rangle^S$ denote convex envelopes associated with the square, bilinear product, cosine, and sine functions, respectively, as defined in~\cite{coffrin2015qc}. Define an impedance parameter $Z_{jk}$ associated with each line $(j,k)\in\mathcal{E}$. The QC relaxation is:
\begin{subequations}
\label{eq:qc}
\begin{align}
& \min \eqref{eq:ac_pf1} ~~ \textrm{s.t.} ~~ \eqref{eq:ac_pf9}\text{--}\eqref{eq:ac_pf15}\\
& \label{eq:qc_pf1} \mathbf{W}_{i i}\in\left\langle v_{i}^{2}\right\rangle^{T}\\
& \label{eq:qc_pf2} \Re\left(\mathbf{W}_{jk}\right)\in\left\langle\left\langle v_{j} v_{k}\right\rangle^{M}\left\langle\cos \left(\theta_{j}-\theta_{k}\right)\right\rangle^{C}\right\rangle^{M}\\
& \label{eq:qc_pf3} \Im\left(\mathbf{W}_{jk}\right)\in\left\langle\left\langle v_{j} v_{k}\right\rangle^{M}\left\langle\sin \left(\theta_{j}-\theta_{k}\right)\right\rangle^{S}\right\rangle^{M}\\
& \label{eq:qc_pf4} S_{jk}+S_{kj}=Z_{jk} l_{jk}\\
& \label{eq:qc_pf5} \left|S_{jk}\right|^{2} \leq \mathbf{W}_{jj} l_{jk}.
\end{align}
\end{subequations}


\subsection{Linear Programming AC (LPAC) Approximation}
The LPAC approximation linearizes the sine function and makes other approximations regarding near-nominal voltage magnitudes to model active and reactive power injections. Similar to the QC relaxation, the LPAC approximation also introduces voltage magnitude and phase angle variables $v_i$ and $\theta_i$ for each bus $i\in\mathcal{N}$ along with lifted variables $\phi_{jk}$ representing $\cos(\theta_j-\theta_k)$ using convex envelopes $\left\langle\,\cdot\right\rangle^{LPAC}$ that are related to the QC relaxation's cosine envelopes $\left\langle\,\cdot\right\rangle^C$. Each line $(j,k) \in\mathcal{E}$ has conductance $g_{jk}$ and susceptance $b_{jk}$. The cold-start version of the LPAC approximation is:
\begin{subequations}
\label{eq:lpac}
\begin{align}
& \min \eqref{eq:ac_pf1} ~~ \textrm{s.t.} ~~ \eqref{eq:ac_pf10}\text{--}\eqref{eq:ac_pf12}\\
& \label{eq:lpac_P} \Re\left(S_{jk}\right) = g_{jk} - (g_{jk} \phi_{jk} + b_{jk}(\theta_j - \theta_k))\\
& \label{eq:lpac_Q} \Im\left(S_{jk}\right) = -b_{jk}(1 + v_j - v_k) - (g_{jk} (\theta_j - \theta_k) - b_{jk} \phi_{jk})\\
& \phi_{jk} \in \left\langle\cos \left(\theta_{j}-\theta_{k}\right)\right\rangle^{LPAC} \\
& \label{eq:lpac_thetadiff} -\overline{\theta}_{jk} \leq \theta_j - \theta_k \leq \overline{\theta}_{jk}.
\end{align}
\end{subequations}

\section{Restoring AC Power Flow Feasibility}
\label{sec:Operating Point Recovery Algorithm}

Each of the formulations in Section~\ref{sec:Preliminaries} have their own advantages and disadvantages. The OPF problem~\eqref{eq:opf} gives a feasible solution if the solver converges, but the problem is nonconvex and NP-Hard, meaning that solvers may fail to converge or converge to a local solution. Moreover, the nonconvexity of this problem imposes significant challenges for solving extensions of~\eqref{eq:opf} that involve discrete and stochastic variables. Conversely, the relaxations and approximations~\eqref{eq:socp1}--\eqref{eq:lpac} are convex but are not guaranteed to give a solution that satisfies the power flow equations. To address this issue, this section develops a method for restoring voltage phasors and power injections that satisfy the AC power flow equations using information from relaxed and approximated solutions.

This section presents a method that, compared to previous methods, uses more information available in the solution by not fixing any variables to specific values in order to achieve better accuracy. As shown in Table~\ref{tab:analogy}, we draw on ideas from state estimation with the voltage phasors, power injections, and line flows from a relaxation or approximation playing a role analogous to noisy measurements. We emphasize that we are \emph{not} suggesting to use measured quantities from the actual system. Rather, similar to the manner by which state estimation algorithms resolve inconsistencies between noisy measurements, our method seeks the voltage phasors that most closely match the voltage phasors, power injections, and line flows resulting from a power flow relaxation or approximation.
\ifarxiv

\else
\fi

Analogous to how state estimation algorithms use the variation associated with sensor noise to weight measured quantities, our method includes weighting parameters associated with the outputs of each quantity from the relaxed or approximated OPF solution. However, rather than being determined by the physical characteristics of a sensor, we can customize our weighting parameters based on the inconsistencies (with regard to the AC power flow equations) between various quantities in the solution to the relaxed or approximated problem. To compute good values for these weighting parameters, we propose a gradient descent based technique. 
\subsection{Restoration Method}

We next propose our method for restoring AC power flow feasible points from the solutions to convex relaxations and approximations. This method seeks the voltage phasors that are as close as possible to the actual OPF solution's voltage phasors based on the voltage phasors, power injections, and line flows from the relaxed or approximated OPF solution.
\begin{table}[t]
\vspace*{0.75em}
\caption{Analogy between the proposed method and SE}
\label{tab:analogy}
\centering
\begin{tabular}{|c|c|} 
 \hline 
 \thead{\textbf{Proposed restoration method}} & \thead{\textbf{State estimation}}\\ 
 \hline\hline
 \thead{Solutions from \\relaxed or approximated models} & \thead{Measurements from \\physical sensors} \\ 
 \hline
 \thead{Inconsistencies in relaxed or \\ approximated solutions} & \thead{Noise from physical sensors} \\ 
 \hline
\thead{Weighting parameters} & \thead{Variance of the \\  measurement noise} \\
\hline
\end{tabular}
\vspace{-1em}
\end{table}

We first introduce notation based on typical presentations of state estimation algorithms to show how we leverage this mathematical machinery, while again emphasizing that we do not use measured quantities from physical sensors. We seek the voltage phasors, denoted as $x$, that are most consistent with the voltage magnitudes, phase angles, power flows, and power injections from the solution to a relaxed or approximated OPF solution, which we gather into a vector $z$. Let $m$ denote the number of these quantities, i.e., the length of $z$, and $n$ denote the number of voltage magnitudes plus the number of (non-slack) voltage angles, i.e., the length of $x$. We relate $x$ and $z$ through the AC power flow model, denoted as $h(x)$:
\begin{equation}
z_i=h_i(x)+e_i, \quad i=1,\ldots, m,
\end{equation}
where $e_i$ indicates the inconsistency between $x$ and $z$ relative to the AC power flow model $h(x)$.


Similar to typical state estimation algorithms, our method uses a weighted least squares formulation:
%


\begin{equation}\label{eq:se_setup}
 \min J(x) = e^{T} \Sigma e,
\end{equation}
\textcolor{black}{where $\Sigma$ is a diagonal weighting matrix. (In a state estimation application, $\Sigma$ would be the covariance matrix for the sensor noise. Conversely, we permit $\Sigma$ to be any diagonal matrix.) We will propose a method for choosing $\Sigma$ in the following subsection.}

Computing~\eqref{eq:se_setup} is accomplished by solving:
\begin{equation}\label{eq:se_g}
    g(x)=\frac{\partial J(x)}{\partial x}=-(H(x))^{T} \Sigma (z-h(x))=0,
\end{equation}
where $H(x)$ is the Jacobian matrix associated with the AC power flow model $h(x)$, i.e., $H(x)=\frac{\partial h(x)}{\partial x}$. Applying a Newton-Raphson method to~\eqref{eq:se_g} yields the $k$-th iterate:
\begin{equation}
\label{eq:newton0}
x^{k+1}=x^{k}-(G(x))^{-1}g(x),
\end{equation}
where $G(x)=\frac{\partial g(x)}{\partial x}$. Thus, the solution to~\eqref{eq:se_setup} is obtained with the following iterative algorithm:
\begin{align}
\nonumber
& x^{k+1}=x^{k}+ \\
\label{eq:newton}& \quad \left((H(x^{k}))^{T} \Sigma \, H(x^{k})\right)^{-1} ((H(x^{k}))^{T} \Sigma \left(\mathrm{z}-h\left(x^{k}\right)\right).
\end{align}

In contrast to existing restoration methods, observe that the value of $x$ depends on all quantities from the relaxed or approximated OPF solution in $z$. 
Thus, we leverage more information available in the relaxed or approximated solution \textcolor{black}{compared to approaches that fix a subset of variables to the values from the solution (e.g., generator active power outputs and voltage magnitudes) and ignore the rest (e.g., generator reactive power outputs, line flows, voltage phase angles, etc.)}. As we will show numerically in Section~\ref{sec:Results and Discussion}, this facilitates more accurate restorations of AC power flow feasibility.


\subsection{Determining the Weighting Parameters}
The weighting parameters $\Sigma$ can have a significant impact on the accuracy of the point obtained from the algorithm~\eqref{eq:newton}. Conceptually, we would like to choose larger values of $\Sigma_{ii}$ for quantities $z_i$ from the relaxed or approximated solution which more accurately represent the true solution to the OPF problem. However, the accuracy of a particular $z_i$ with respect to the true OPF solution is unknown a priori and difficult to estimate. Accordingly, 
we employ a method inspired by algorithms for training machine learning models in order to determine the weighting parameters.
This method first solves a randomly generated representative set of OPF problems along with the relaxation or approximation of interest to create a training dataset. The method then uses a gradient descent method that iterates between solving the proposed restoration method and computing steps which minimize a loss function based on the difference between the restored point and the true OPF solution across the training dataset.

The sensitivities of the restored point $x$ with respect to the weighting parameters $\Sigma$ are key to this gradient descent method. These sensitivities $\frac{ \partial x_{R}}{\partial \Sigma}$ are:
\begin{align}\nonumber
  & \frac {\textrm{vec}(\partial x_{R})}{\textrm{vec}( \partial \Sigma)}  = \bigg((z-h) - (H(H^T \Sigma H )^{-1}\\ & \label{vector} \quad\qquad H^T \Sigma (z-h)\bigg) 
  \otimes \bigg((H^T \Sigma H)^{-1} H^T \bigg)^T,
\end{align}
where $\otimes$ is the Kronecker product and $\textrm{vec}(\,\cdot\,)$ denotes the vectorization of a matrix. (Note that the dependencies on $x$ are suppressed for notational brevity in~\eqref{vector}.) The vectors $z$ and $h(x)$ have length $m$ and the matrix $H$ has size $m\times n$. The resulting sensitivities are described by a $n \times m^2$ matrix. 
\ifarxiv
The appendix provides the mathematical derivation for \eqref{vector}.
\fi

Using these sensitivities, we seek weighting parameters that minimize a loss function formulated as the squared difference between the OPF solutions, whose voltage magnitudes and angles are denoted as $x_{AC}$, and points from our restoration method, whose voltage magnitudes and angles are denoted as $x_{R}$, across the set of training data. Thus, \textcolor{black}{following typical practice when training machine learning models, we formulate an two-norm loss function}:
\begin{equation}
  \label{eq:objective2} F(\Sigma)=\frac{1}{2} \Psi^{T}(\Sigma)\Psi(\Sigma), 
\end{equation}
where $\Psi(\Sigma)=x_{R}(\Sigma)-x_{AC}$. 

To minimize this loss function, we use a gradient descent algorithm. The gradient of the objective with respect to the weighting parameters is denoted as~$q_t$:
\begin{equation}\label{eq:gradient}
q_t=\nabla_\Sigma F(\Sigma)=J^{T}_{\Psi}\Psi=\frac{ \partial x_{R}}{\partial \Sigma}(x_{R}-x_{AC}).  
\end{equation}

Using this gradient, one can find the optimal weighting parameters $\Sigma$ using an iterative method. There are many variants of gradient descent algorithms, such as batch gradient, momentum, AdaGrad, Adam, improved Adam, etc., each of which has their own advantages and disadvantages. 
We use the Adam algorithm since we empirically found it to perform best for this application. The Adam algorithm is commonly used for training machine learning models and involves the following steps~\cite{kingma2014adam}:
\begin{subequations}
\label{eq:adam}
\begin{align}
 m_t &= \beta_1 m_{t-1} + (1-\beta_1)q_t, \\
 v_t &= \beta_2 v_{t-1} + (1-\beta_2)q_t^2,\\
 \hat{m}_t &= \frac{m_t}{1-\beta_1^t}, \\
 Q_{t} &= \frac{v_t}{1-\beta_2^t},\\
 \Sigma_{t} &= \Sigma_{t-1} - \eta \frac{\hat{m}_t}{\sqrt{Q_{t}} + \epsilon},
\end{align}
\end{subequations}
where $m_t$ and $v_t$ are the first and second moments of the gradients at time step $t$, $\eta$ is a learning rate (step size), $q_t$ is the gradient at time step $t$, and $\beta_1$ and $\beta_2$ are exponentially decaying hyperparameters for the first and second moments, respectively.
%
Using the gradient descent method shown in Algorithm~1, the optimal weighting parameters are obtained iteratively over the training dataset. We focus on a diagonal matrix $\Sigma$, as in traditional state estimation, with extensions to more general matrices being the subject of future work.
The training dataset is created by randomly sampling different loading conditions within the forecast range of demands expected over the next time period (e.g., the variation predicted within the next day's load forecast). After the off-line execution of Algorithm~1, the resulting weights are applied to restore AC power flow feasibility for solutions to OPF relaxations and approximations computed on-line.

     
      

  



\begin{algorithm}[t]

\DontPrintSemicolon
\caption{Computing Weight Parameters}  

  \KwInput{$\eta,\epsilon, \beta_1,\beta_2$: Gradient descent parameters


\hspace{1.1cm}$\Sigma^{init}$: Initial weighting parameters
}
  \KwOutput{$\Sigma^{opt}$: Optimal weighting parameters}
  \KwData{OPF problem data}
  Generate loading scenarios $s$ and store in the set~$\mathcal{S}$
  
  Solve OPF problems~\eqref{eq:opf} for each scenario $s\in\mathcal{S}$ and store the results in $x_{AC}$ 
  
  Solve relaxed or approximated OPF problems for each scenario $s\in\mathcal{S}$ and store the results in $x_{R}$
  


   $\Sigma^{1} \leftarrow \Sigma^{init} $ 
   
  \While{\begin{math} iter \leq max\_iter \end{math}}
   {
        \begin{math}
        \nabla F(\Sigma)=0
        \end{math}
        
    	\For{$s \in \, \mathcal{S}$}    
    	{ 
            
                 
            
        	        Run restoration method~\eqref{eq:newton} to an accuracy of $\varepsilon$ and store solution in $x_{R}$
        	   
        	        
        	        
        	        $\nabla F(\Sigma) \leftarrow \nabla F(\Sigma)+ \frac{ \partial x_{R}}{\partial \Sigma} (x_{R}-x_{AC})$
        	     
        	
        	   Store solutions: $X_{R}  \leftarrow  x_{R}$
        	    
        }
        \begin{math}
        \Delta \Sigma =-\eta \frac{\hat m_{t}}{\sqrt{Q_{t}}+\epsilon}
        \end{math}
        
        \begin{math}
        \Sigma_{t}=\Sigma_{t-1}+ \Delta \Sigma
        \end{math}
        
        \begin{math}
        \mathcal{L}=\frac{1}{2N-1}\sum_{s \in \mathcal{S}}\sum_{i \in \mathcal{N}} \left(||X_{R}^{i,s}-X_{AC}^{i,s}||_2^2\right)
        \end{math}
        \begin{math}
        iter\leftarrow iter+1
        \end{math}
   }

  $\Sigma^{opt} \leftarrow \Sigma$

 \label{alg:algorithm1}

\end{algorithm}



  
       
       

            
                 
            
        	   
        	        
        	    


\section{Experimental Results and Discussion}
\label{sec:Results and Discussion}
This section presents numerical results for evaluating the performance of the proposed restoration algorithm for several convex relaxations and approximations.
\ifarxiv
\subsection{Experiment Setup}
\else
\fi
We evaluate the performance of proposed restoration method by applying the SOCP, SDP, and QC relaxations and LPAC approximation to the PJM five-bus, IEEE 14-bus, IEEE 57-bus, and IEEE 118-bus systems~\cite{pglib}. Datasets of 10,000 scenarios for training and testing purposes were generated by multiplicatively perturbing the nominal loads in these test cases by a normal random variable with zero mean and standard deviation of 0.1. The OPF problems as well as the relaxations and approximations (e.g., $x_{AC}$) were computed using \texttt{PowerModels.jl} \cite{coffrin2018powermodels} with the solvers Ipopt~\cite{wachter2006implementation} and Mosek%
\ifarxiv
~\cite{mosek}
\else
\fi
~on an Apple laptop with a 10-core M1 Pro CPU and 32~GB of RAM. The restoration method was implemented in Python 3.0 using a Jupyter notebook.

\ifarxiv
\subsection{Benchmarking Approach}
\fi
To assess the performance of various restoration methods, we consider the two-norm of the difference between the OPF solution's voltage phasors and the restored solution's voltage phasors across all samples, denoted as $\mathcal{L}$. For instance, when considering the QC relaxation:
\begin{equation}
\label{eq:lossfunction2}
    \mathcal{L}=\frac{1}{2N-1}\sum_{s \in \mathcal{S}}\sum_{i \in \mathcal{N}}\left(||v_{QC}^{i,s}-v_{AC}^{i,s}||_2^2+||\theta_{QC}^{i,s}-\theta_{AC}^{i,s}||_2^2\right),
\end{equation}
with voltage magnitudes in per unit and angles in radians.

We consider four restoration methods. The first simply compares the voltage magnitudes and angles from the relaxed or approximated solution directly (without any processing) to the OPF solution. Note that this method typically does not yield an AC power flow feasible point and is thus not suitable for many practical applications. Also note that this method is inapplicable to the SDP and SOCP relaxations which lack variables corresponding to the voltage phase angles. The second method, denoted as ``benchmark'' in the discussion below, solves the power flow problem that results from fixing the voltage magnitudes at all generator buses and the active power injections at non-slack generator buses to the output of the relaxation of approximation, as discussed in~\cite{venzke2020inexact}. The third is our proposed restoration method with the weighting parameters set to (heuristically determined) initial values of $\Sigma$ that weight the voltage magnitudes and phase angles \textcolor{black}{with values of $10^{4}$ and the power injections and line flows with values of $10^{3}$}. The fourth is our restoration method with the weighting parameters computed using Algorithm~1, trained using 8,000 test scenarios. We use 2,000 test scenarios to evaluate each method.


\subsubsection{Training the Weighting Parameters}

\begin{figure}[t]
    \vspace*{0.75em}
    \centering
    \includegraphics[width=8cm]{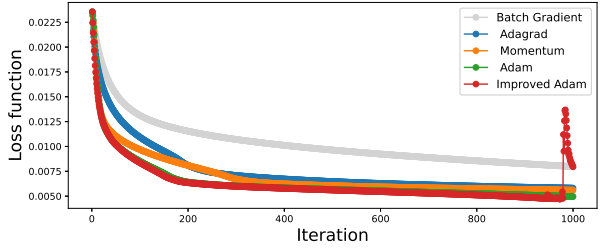}
     \caption{The loss function during the training process using various gradient descent algorithms for the 5-bus system.}
     \label{fig:loss_function}
     \vspace*{-1em}
 \end{figure}

\ifarxiv
To compare the performance of different gradient descent algorithms for computing the weighting parameters, we ran a sample loss function using the batch gradient, momentum base gradient, Adagrad, Adam, and improved Adam methods with $20$ training scenarios for $1,000$ iterations with the heuristically determined initialization described above. The results are shown in Fig.~\ref{fig:loss_function} for the five-bus test system. While each of these methods reduces the loss function, the Adam method outperforms the other methods. 

\else\fi

\begin{figure}[t]
    \vspace*{0.75em}
    \centering
    \includegraphics[width=2cm,angle=270]{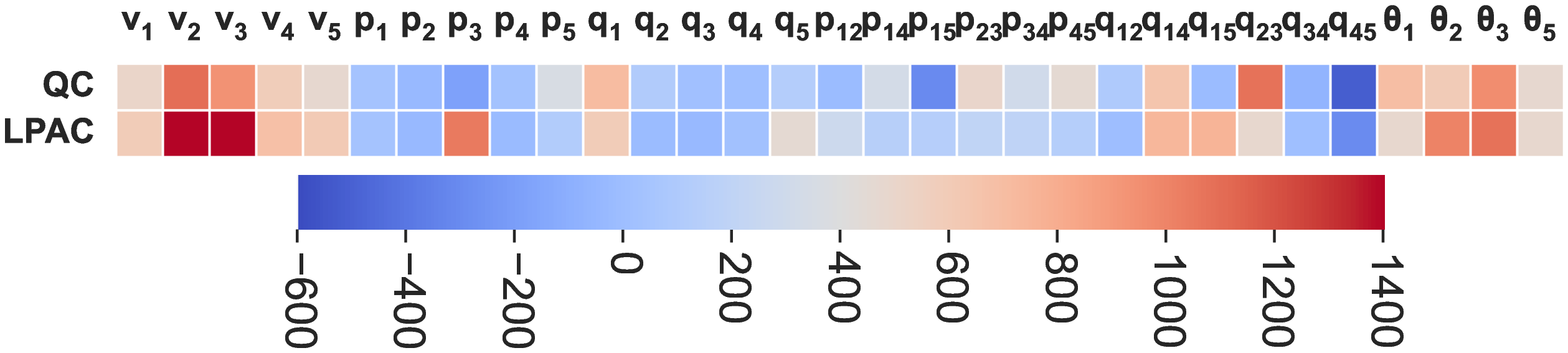}
    \includegraphics[width=2cm,angle=270]{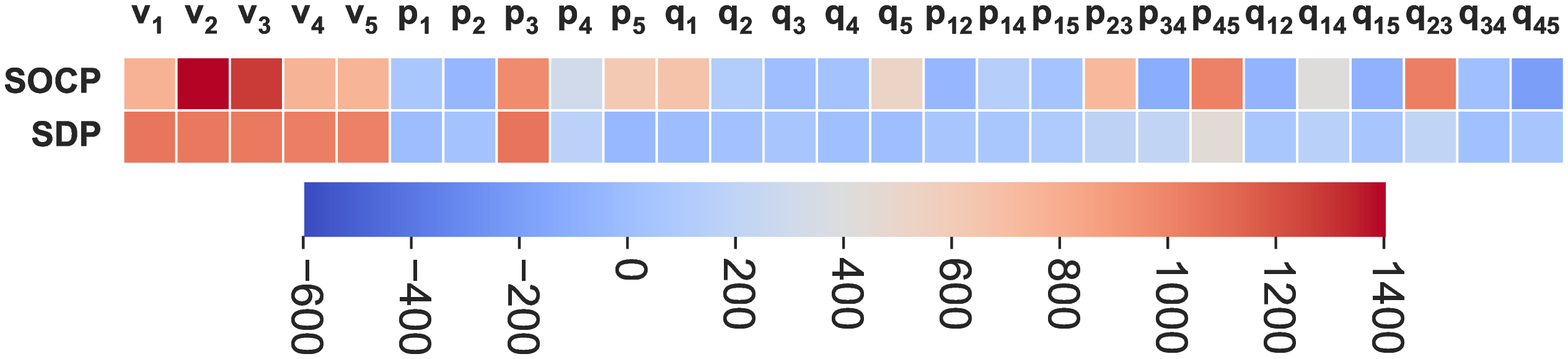}
     \caption{Trained diagonal elements of weight matrices for QC, SOCP, SDP, and LPAC in the 5-bus system.}
     \label{fig:weight_matrix}
    \vspace*{-0.5em} 
 \end{figure}

The weighting parameters (i.e., the diagonal elements of $\Sigma$) resulting from Algorithm~1 for the five-bus system with the QC, SOCP, and SDP relaxations and the LPAC approximation are visualized in Fig.~\ref{fig:weight_matrix}. Observe that some quantities are assigned much larger weights than others. For instance, the algorithm puts more weight on voltage magnitudes and angles, especially at buses~2 and~3, for this test case, which shows that these quantities are more valuable for predicting the actual OPF solution.


\begin{figure}
\centering
\begin{subfigure}{0.44\textwidth}
    \includegraphics[width=\textwidth]{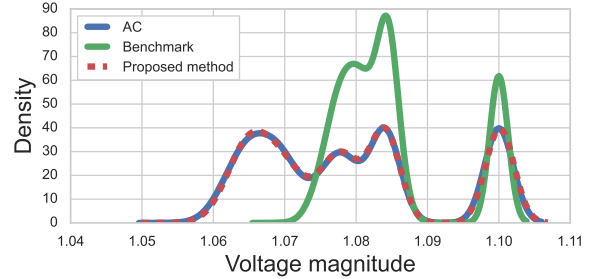}
    \caption{Voltage magnitudes}
    \label{fig:first}
\end{subfigure}
\hfill
\begin{subfigure}{0.44\textwidth}
    \includegraphics[width=\textwidth]{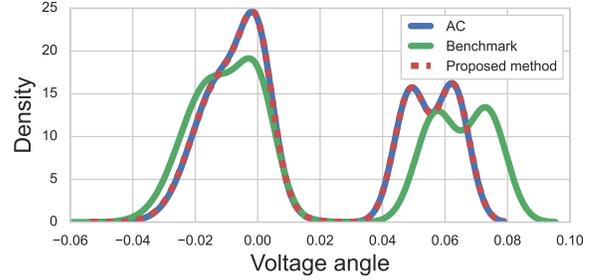}
    \caption{Voltage angles}
    \label{fig:second}
\end{subfigure}
        
\caption{Comparison of the voltages for the AC solution vs. the benchmark and proposed methods with the optimized weighting parameters for the SDP relaxation in 5-bus system.}
\label{fig:SE_W_voltage_phasors}
     \vspace*{-1em}
\end{figure}


\subsubsection{Performance Evaluation}

\begin{figure}[t]
    \centering
    \vspace*{0.75em}
    \includegraphics[width=8cm]{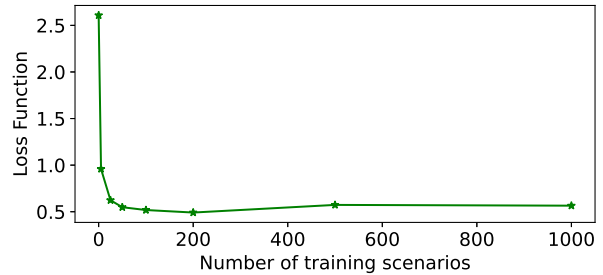}
     \caption{The loss function with different training scenarios.}
     \label{fig:loss_vs_samples}
 \end{figure}

We next show the effectiveness of the proposed restoration method using the test datasets that were unseen during the weighting parameter calculation in Algorithm~1 (i.e., the 2,000 test scenarios). The values of the loss function~\eqref{eq:lossfunction2} for each solution recovery method are given in Table~\ref{table:loss_function}. To further illustrate the performance of the proposed method, the voltage phasors from OPF solutions, benchmark, and the restored point from the proposed method (with optimized weighting parameters) are compared in Fig.~\ref{fig:SE_W_voltage_phasors} for the SDP relaxation. 
Observe that the proposed method's solution closely matches the actual OPF solution.
\ifarxiv

\else
\fi
As can be seen from both Table~\ref{table:loss_function} and Fig.~\ref{fig:SE_W_voltage_phasors}, the proposed restoration method obtains high-quality AC power flow feasible points from the solutions to relaxed and approximated OPF problems. The loss functions from the proposed method are several orders of magnitude smaller than the other approaches, including the benchmark approach, \textcolor{black}{except for the IEEE 14-bus system where the SDP relaxation provides the globally optimal OPF solution}.\footnote{\textcolor{black}{Assessing whether the SDP relaxation provides the globally optimal solution to an OPF problem involves a straightforward rank computation that could be done prior to executing our proposed recovery method~\cite{lavaei2011zero}.}} Moreover, the optimized weighting parameters significantly improve the loss function relative to the heuristically determined parameters. 
\ifarxiv
The loss functions when computed with the heuristically determined weighting parameters for the IEEE 14-bus system in comparison with the optimized weighting parameters are $91$, $1581$, $1054$, and $1093$ times larger for LPAC, QC, SOCP, and SDP formulations, respectively.
\else
\fi
\ifarxiv

\else
\fi
Furthermore, to show how the proposed method's performance scales with different numbers of training scenarios, we optimized the weighting parameters using varying number of scenarios for the five-bus test system with the QC relaxation. Fig.~\ref{fig:loss_vs_samples} shows the obtained loss function on the test dataset for varying numbers of training scenarios. With 200 training scenarios for this test case, the restoration method can efficiently learn the weighting parameters. The results show little improvement (and even some minor degradation) when using more scenarios. 
\ifarxiv

\else
\fi

We additionally show the execution time per scenario with optimized weights in Table~\ref{table:comput}. 
We note that the main computational burden in this problem is the training of the weighting parameters, which is done off-line. In this off-line training, solving the relaxed and approximated OPFs and running the restoration method to compute the weights are the most time-consuming steps in the algorithm. The on-line execution is fast (comparable to a power flow evaluation), as shown by the computation times in Table~\ref{table:comput}.
\ifarxiv
Our future work aims to assess variants such as stochastic gradient descent that may lead to substantial computational improvements during the off-line computations of the weighting matrix.
\else
\fi
\newcommand{\ra}[1]{\renewcommand{\arraystretch}{#1}}
\begin{table}
\tiny
\vspace*{0.75em}
\centering
\caption{Loss Function Evaluated for Different Test Cases Using Various Relaxations and Approximations}
\ra{1.1}
\begin{tabularx}{\columnwidth}{@{}cccccc@{}}
\cmidrule{1-6}
& &\multicolumn{4}{c}{\scriptsize Convex Relaxations and Approximation} \\
\cmidrule{3-6}
\scriptsize Test Case & \scriptsize Method & \scriptsize QC&\scriptsize SOCP &\scriptsize SDP &\scriptsize LPAC \\ \cmidrule{1-6}
\multirow{4}{*}{\scriptsize PJM 5-Bus} &
\scriptsize R/A  sol. &\scriptsize  0.6709 &\scriptsize --- & --- &\scriptsize 0.4996 \\
&\scriptsize Benchmark &\scriptsize 0.6069 &\scriptsize 0.6077 &\scriptsize 0.1279 &\scriptsize0.4748\\
&\scriptsize SE with $\Sigma^{init}$ &\scriptsize 0.2886 &\scriptsize0.2355 &\scriptsize 1.0840 &\scriptsize 0.2697 \\
&\scriptsize SE with $\Sigma^{opt}$ &\scriptsize \textbf{0.0201} &\scriptsize \textbf{0.0206} &\scriptsize \textbf{0.0002} &\scriptsize \textbf{0.0322} \\

\cline{1-6}
\multirow{4}{*}{\scriptsize IEEE 14-Bus} &
\scriptsize R/A  sol.&\scriptsize  3.7926  & ---    &\scriptsize ---     &\scriptsize  0.5655  \\
&\scriptsize Benchmark                     &\scriptsize  0.0009 &\scriptsize 0.0010&\scriptsize\hphantom{1} \textbf{0.000003}  &\scriptsize 0.1937\\
&\scriptsize SE with $\Sigma^{init}$      &\scriptsize  0.2284 &\scriptsize 0.2457 &\scriptsize 0.2540  &\scriptsize 0.6110 \\
&\scriptsize SE with $\Sigma^{opt}$       &\scriptsize  \textbf{0.0001}&\scriptsize \textbf{0.0002} &\scriptsize 0.0002 &\scriptsize\textbf{0.0067}\\
\cline{1-6}
\multirow{4}{*}{\scriptsize IEEE 57-Bus} &\scriptsize
 R/A sol. &\scriptsize  1.8558 &\scriptsize --- &\scriptsize ---  &\scriptsize 0.5538   \\
&\scriptsize Benchmark                      &\scriptsize  0.0567 &\scriptsize 0.0566 &\scriptsize  0.0463 &\scriptsize 0.7968 \\
&\scriptsize SE with $\Sigma^{init}$        &\scriptsize1.3544 &\scriptsize 1.4155&\scriptsize  1.0713 &\scriptsize  2.4205\\
&\scriptsize SE with $\Sigma^{opt}$        &\scriptsize \textbf{0.0565}&\scriptsize \textbf{0.0558} &\scriptsize \textbf{0.0320} &\scriptsize\textbf{0.1201}\\

\cline{1-6}
    \multirow{4}{*}{ \scriptsize IEEE 118-Bus} &\scriptsize
R/A  sol. &\scriptsize  6.1651 &\scriptsize --- &\scriptsize ---  &\scriptsize 4.8066 \\
&\scriptsize Benchmark                       &\scriptsize 0.2051  &\scriptsize 0.2056 &\scriptsize  0.0113 &\scriptsize  5.0810\\
&\scriptsize SE with $\Sigma^{init}$        &\scriptsize 5.2822 &\scriptsize 7.3201 &\scriptsize 6.8255  &\scriptsize 4.5119\\
&\scriptsize SE with $\Sigma^{opt}$         &\scriptsize\textbf{0.0313} &\scriptsize \textbf{0.0325} &\scriptsize \textbf{0.0111} &\scriptsize\textbf{0.1463}\\

\cmidrule{1-6}
\end{tabularx}
\begin{tablenotes}
 \item[*] \scriptsize The best performing method (i.e., smallest loss function) is bolded for each test case.
 \item[*] \scriptsize ``R/A sol.'' indicates the solution to the relaxation or approximation.
\end{tablenotes}
\label{table:loss_function}
\end{table}

\begin{table}\centering
 \vspace*{0.70em}
\caption{Average Execution Time per Scenario (sec)}
\ra{1.1}
\begin{tabular}{cccccc}
\toprule
&\multicolumn{4}{c}{Convex Relaxations and Approximation} \\
\cmidrule{2-5} 
Case &QC& SOCP & SDP & LPAC \\ \midrule
\multirow{1}{*}{PJM 5-Bus} &
 0.005 & 0.004 & 0.004 &0.006\\
 
\hline
\multirow{1}{*}{IEEE 14-Bus} &
0.013 & 0.021 &0.025 & 0.020  \\

\hline
\multirow{1}{*}{IEEE 57-Bus} &
  0.034 &0.030 & 0.033 & 0.042\\

\hline
\multirow{1}{*}{IEEE 118-Bus} &
0.159  & 0.392 & 0.338 &  0.161  \\

\bottomrule
\end{tabular}
\label{table:comput}
\end{table}





\section{Conclusion}
\label{sec:Conclusion}
\ifarxiv
To recover AC power flow feasible points from the solutions to relaxed and approximated OPF problems, this paper proposes a solution restoration method inspired by state estimation techniques. Treating quantities from the solutions of relaxed or approximated OPF problems (i.e., voltage phasors, power injections, and line flows) analogous to measurements from physical sensors in a state estimation algorithm, the proposed method finds an operating point that is feasible with respect to the AC power flow equations. Moreover, we improve this method by adjusting weighting parameters using a gradient descent algorithm. After computing the weighting parameters based on training datasets, application of the method to the unseen test datasets demonstrates that the solution restoration method can achieve several orders of magnitude improvement in accuracy relative to existing methods. 
\fi

Our future work includes 1)~improving computational tractability of the off-line training calculations using, e.g., stochastic gradient descent techniques, \textcolor{black}{2)~extending this work to consider restoration of AC power flow feasibility for the outputs of machine learning algorithms such as~\cite{pan2022deepopf, chatzos2022, Zamzam2020, kody2022}}, \textcolor{black}{3)~incorporating more information from the relaxed or approximated solutions using, e.g., combination of different relaxations and approximations,} 
and 4)~using the optimal weights to characterize the behavior of power flow relaxations and approximations.

\ifarxiv

\vspace{-1em}
\appendix

The sensitivity of the voltage phasors obtained from the state estimation inspired algorithm (i.e., $x_{R}$) with respect to the weight matrix (i.e., $\Sigma$) is computed by~\eqref{vector}, which is derived as follows:
\begingroup\makeatletter\def\f@size{10}\check@mathfonts
\begin{subequations}
\begin{align}
  &\begin{aligned}
   \label{eq:app1}
     Y=\left(H^{T} \Sigma H\right)^{-1} H^{T} \Sigma\left(\mathrm{z}-h\left(x^{k}\right)\right)
  \end{aligned}
    \end{align} 
 \begin{align}
  &\begin{aligned}
  \label{eq:app2}
\begin{split}
dY=
\overbrace{d\bigg(\left(H^{T} \Sigma H\right)^{-1} H^{T} \bigg) \Sigma\left(\mathrm{z}-h\left(x^{k}\right)\right)}^{I} \\
+ 
\underbrace{\left(H^{T} \Sigma H\right)^{-1} H^{T} d\bigg(\Sigma\left(\mathrm{z}-h\left(x^{k}\right)\right)\bigg)}_{U}
\end{split}
  \end{aligned}
      \end{align} 
 \begin{align}
 &\begin{aligned}
      \label{eq:app3}
     \begin{split}
        U=\left(H^{T} \Sigma H\right)^{-1} H^{T} d\Sigma\left(\mathrm{z}-h\left(x^{k}\right)\right)
\end{split} 
 \end{aligned}
     \end{align}
\vspace*{-0.5em}
 \begin{align}
  &\begin{aligned}
      \label{eq:app4}
\begin{split}
I=d\bigg(\left(H^{T} \Sigma H\right)^{-1}\bigg) H^{T}  \Sigma\left(\mathrm{z}-h\left(x^{k}\right)\right)
\end{split}
  \end{aligned}
      \end{align} 
 \begin{align}
  &\begin{aligned}
   \label{eq:app5}
\begin{split}
I=\bigg(-\left(H^{T} \Sigma H\right)^{-1} d\left (H^{T} \Sigma H\right) \left(H^{T} \Sigma H\right)^{-1}  \bigg) \\
H^{T}  \Sigma\left(\mathrm{z}-h\left(x^{k}\right)\right)
\end{split}
  \end{aligned}
      \end{align} 
 \begin{align}
  &\begin{aligned}
   \label{eq:app6}
\begin{split}
I=\bigg(-\left(H^{T} \Sigma H\right)^{-1} H^{T} d\Sigma H  
\left(H^{T} \Sigma H\right)^{-1}  \bigg) \\
H^{T}  \Sigma\left(\mathrm{z}-h\left(x^{k}\right)\right)
\end{split}
  \end{aligned}
  \end{align} 
 \begin{align}
  &\begin{aligned}
   \label{eq:app7}
\begin{split}
dY=-\left(H^{T} \Sigma H\right)^{-1} H^{T} d\Sigma H \left(H^{T} \Sigma H\right)^{-1}
H^{T} \Sigma\\\left(\mathrm{z}-h\left(x^{k}\right)\right)+ \left(H^{T} \Sigma H\right)^{-1} H^{T} d\Sigma\left(\mathrm{z}-h\left(x^{k}\right)\right)
\end{split}
  \end{aligned}
      \end{align} 
 \begin{align}
  &\begin{aligned}
   \label{eq:app8}
\begin{split}
\textrm{vec}(dY)=\textrm{vec} \bigg [-\left(H^{T} \Sigma H\right)^{-1} H^{T} d\Sigma H \\ \left(H^{T} \Sigma H\right)^{-1}
H^{T} 
\Sigma\left(\mathrm{z}-h\left(x^{k}\right)\right) \bigg] \\
+ \textrm{vec} \bigg[ \left(H^{T} \Sigma H\right)^{-1} H^{T}  d\Sigma
\left(\mathrm{z}-h\left(x^{k}\right)\right) \bigg]
\end{split}
  \end{aligned}
\end{align} 
 \begin{align}
  &\begin{aligned}
   \label{eq:app9}
     \begin{split}
      \textrm{vec}(dY)= \bigg [ - \bigg(H \left(H^{T} \Sigma H\right)^{-1} H^{T} \Sigma 
     \left(\mathrm{z}-h\left(x^{k}\right)\right) \bigg) ^{T} \\
      \otimes \bigg( \left(H^{T} \Sigma H\right)^{-1} H^{T} \bigg) \bigg] \textrm{vec} (d\Sigma) \\
       +\bigg[  (z-h)^{T} \otimes \bigg( \left(H^{T} \Sigma H\right)^{-1} H^{T} \bigg)  \bigg] \textrm{vec} (d\Sigma)
\end{split}
  \end{aligned}\\
  &\begin{aligned}
   \label{eq:app10}
\begin{split}
     \textrm{vec}(dY)= \bigg [(z-h)^{T} \otimes \bigg( \left(H^{T} \Sigma H\right)^{-1} H^{T} \bigg)\\ - \bigg(H \left(H^{T} \Sigma H\right)^{-1} H^{T} \Sigma\left(\mathrm{z}-h\left(x^{k}\right)\right) \bigg) ^{T}  \\
      \otimes \bigg( \left(H^{T} \Sigma H\right)^{-1} H^{T} \bigg) \bigg] \textrm{vec} (d\Sigma) 
\end{split}
  \end{aligned}
      \end{align} 
 \begin{align}
  &\begin{aligned}
  \label{eq:app11}
   \begin{split}
       \frac {\textrm{vec}(\partial Y)}{\textrm{vec}( \partial \Sigma)}  = \bigg ((z-h) - (H(H^T \Sigma H )^{-1} H^T \\ \Sigma (z-h)\bigg)
       \otimes \bigg ((H^T \Sigma H)^{-1} H^T \bigg)^T  
\end{split}
 \end{aligned}
\end{align}
\end{subequations}

\endgroup

\else
\fi


\bibliographystyle{IEEEtran}
\ifarxiv
\fi
\bibliography{refs}



\end{document}